# Formation of Highly Tunable Periodic Plasmonic Structures on Gold Films Using Direct Laser Writing

*Kernius Vilkevičius[1]\*, George D. Tsibidis[2], Algirdas Selskis[3], Emmanuel Stratakis[2,4], and Evaldas Stankevičius[1]*

[1]Department of Laser Technologies, Center for Physical Sciences and Technology, Savanoriu Ave. 231, Vilnius, 02300, Lithuania
[2]Institute of Electronic Structure and Laser (IESL), Foundation for Research and Technology (FORTH), Vassilika Vouton, 70013, Heraklion, Crete, Greece
[3]Department of Characterization of Materials Center for Physical Sciences and Technology, Sauletekio Ave. 3, Vilnius, 10257, Lithuania
[4]Department of Physics, University of Crete, 71003, Heraklion, Greece

Corresponding author: K. Vilkevičius: kernius.vilkevicius@ftmc.lt

**Abstract.** Direct laser writing method is a promising technique for the large-scale and cost-effective fabrication of periodic nanostructure arrays exciting hybrid lattice plasmons. This type of electromagnetic mode manifests a narrow and deep resonance peak with a high dispersion whose precise controllability is crucial for practical applications in photonic devices. Here, the formation of differently shaped gold nanostructures using the direct laser writing method on Au layers of different thicknesses is presented. The resonance peak is demonstrated to be highly dependent on the shape of the structures in the array, thus its position in the spectra, as well as the quality, can be additionally modulated by changing the morphology. The shape of the structure and the resonance itself pertain not only on the laser pulse energy but also on the grating period. This overlapping effect occurring at distances smaller than the diameter of the focused laser beam is studied in detail. By taking advantage of the highly controllable plasmonic resonance, the fabricated gratings open up new opportunities for applications in sensing.

## 1. Introduction

Plasmonic materials have gained significant attention due to their resonant properties and applicability in optical components and sensors [1-3], Raman spectroscopy [4, 5], emitter-based devices [6, 7], and solar cells [8]. The driving force of such properties is the excitation of surface plasmon polaritons (SPP) and localized surface plasmons (LSP) in the noble metal. LSPs are excited in metallic particles with sizes smaller than the wavelength, while SPPs are excited on the plain metal surface by indirect coupling of photons with SPPs [9]. Although the most widely used method for such coupling is the Kretschmann configuration, which utilizes a glass prism [10, 11], a simpler and more efficient coupling technique is based on diffraction by fabricating periodic structures on the metal surface [12, 13] when an additional element such as a prism is not required.

The fast fabrication and easy tuning of the absorption peak of such structures are crucial for extensive research and diverse applications. While ion or electron lithography methods are suitable for producing precise nanostructures, they prove to be expensive and time-consuming because these are multi-step fabrication techniques requiring a high vacuum [14-16]. Structures fabricated through chemical synthesis are more uniform and easier to produce on a large scale [17, 18]. However, this method is also multi-step and requires additional chemicals, which increases manufacturing complexity and cost. Therefore, the use of ultrafast lasers is a promising cost- and time-efficient single-step alternative for nanostructuring. Femtosecond lasers enable fast and smooth processing with minimal thermal impact, leading to the fabrication of wavelength-sized structures on the surface of the material [19, 20] or within its volume [21-23]. The large-scale arrays of uniformly shaped structures can be fabricated by direct laser writing (DLW) [24-26] or laser interference lithography (LIL) [27, 28]. LIL is a faster structuring process than the DLW method due to the single exposure formation. However, LIL is limited by the size of the array, and the structures in the entire treated area may be non-uniform due to the distribution of the Gaussian beam in the interfered area. In addition, in the case of LIL, beam alignment can be complicated and time-consuming. DLW involves modifying or removing a material with a directly focused laser beam. This method is attractive because of its high accuracy, flexibility, and simplicity compared to other structuring methods. The morphology of the nanostructures plays a pivotal role in determining the optical [29, 30] or chemical [31, 32] properties of the material. Therefore, the ability to control the size and shape of the particles is crucial. Laser parameters, such as energy [24, 33], or laser wavelength [34, 35], can be adjusted to obtain differently shaped and sized structures. When thin films are irradiated with single pulses of a femtosecond laser, the fast phase transition is initiated and complex nanostructures with energy-dependent morphology at the metal surface are produced [36, 37]. The formation of such hollow nanostructures has been recently widely discussed, and several formation mechanisms have been described. The formation of such single-pulse structures is determined by the thermoelastic stress and induced counterpressure of the substrate leading to the delamination of the film. The surface tension (if a film is not melted) and capillary forces (if a film is melted) decelerate the growth of the structure with the radial flow of the mass along the film with the direction to its axis [38, 39]. Bumps having the shape of a hemisphere are the least pulse energy-requiring structures that form due to induced thermal stresses when the metal layer is heated, causing it to slightly lift upwards. When higher energies are applied and the layer is melted, cones and jets might be obtained as the molten metal gains momentum normal to the surface. During the formation of these shapes, the melt gains velocities towards the structure center, leading to the decrease of the film thickness at the structure sides. Additionally, the counter-motion melt colliding at the center gives rise to jets, both oriented upward and downward (counterjet), the latter being inside the structure [40-42]. The jets are produced due to the hydrodynamic movement of the molten metal and its height is limited by the surface tension caused by the Rayleigh-Plateau instability, causing the break of the jet and ejection of the spherical nanoparticles at even higher energies [43-45].

Periodic arrangement of ordered structures has reached great interest compared to single or randomly distributed structures as the high-quality plasmonic resonances of hybrid modes are excited [46-48]. Arrays of gold bumps, fabricated by a single direct laser pulse on a thin Au layer, with a period comparable to the light wavelength, excite a hybrid plasmonic mode called hybrid lattice plasmon (HLP). This mode is achieved due to the hybridization of SPPs excited on the metal surface, LSPs excited on individual structures, and the scattered light from the grating, while the properties of such gratings are determined by the excitation of Bragg modes [49, 50]. Gold bumps have demonstrated great results having both narrow and deep HLP resonance with a high dispersion relation in the VIS-NIR range [51]. As the morphologies of such structures have already been extensively described, the impact and formation of these in varying thickness metal layers, as well as the period-induced shape change, have not yet been investigated.

Here, the fabrication of different morphology periodical gold nanostructures on varying Au thicknesses is presented. The thin films from 25 nm up to 100 nm thick were observed to be optimal and sustain the highest-quality plasmon resonances having the largest figure of merits (FOMs) [52-54], thus this range was chosen for deeper investigation in this research. The resulting arrays excite HLP, with its spectral characteristics (quality and position in the spectra) being highly dependent on the shape. In this paper, to the best of our knowledge, for the first time, we extensively examine the nanostructure morphology change induced by the period reduction below the overlapping threshold. The change in the period and metal thickness provides an additional degree of freedom to easily tune the morphology of the nanostructures, with precise control of the plasmonic properties.

## 2. Results and Discussion

### 2.1. Formation of periodic nanostructures on Au films of different thickness

The periodic nanostructure arrays were produced on a thin Au film that was deposited onto a clear silica glass using a magnetron sputter-coater. The layer thickness was chosen to be 25 nm, 50 nm, 75 nm, and 100 nm and the deposition was done at a constant 0.27 nm s$^{-1}$ coating rate. The films were prepared without an additional adhesion layer, which not only improves the adhesion properties of the coating but also affects the formation parameters of the structures [55]. Subsequently, the structures were fabricated using a Yb:KGW based femtosecond laser (Pharos, Light Conversion Ltd.) and its second harmonic (515 nm), as the formation was achieved with a single laser pulse (Figure 1a).

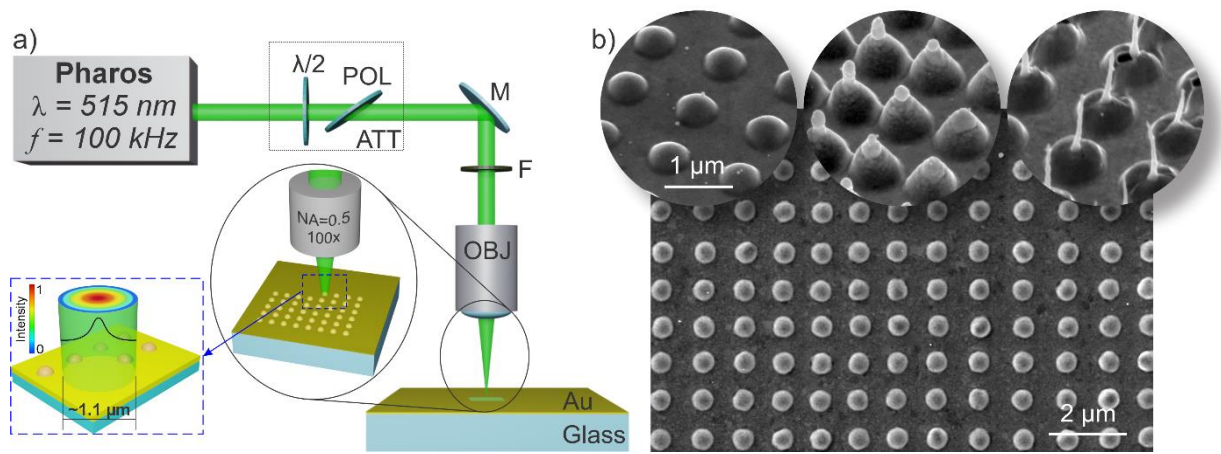

**Figure 1.** a) Nanostructure formation scheme: fs-laser Pharos, attenuator (ATT) used for the energy control, mirror (M), filter (F) used for an additional energy reduction, and focusing objective (OBJ). The beam is tightly focused onto the metal surface using a high-NA objective. Due to the Gaussian beam distribution, the structures are fabricated only in the central part of the beam. b) SEM images of the top-view periodic array and different morphological states (bumps, cones, jets) at a 53° angle.

The DLW technique was employed to produce large-scale arrays (3×3 mm$^2$). The fabrication time of such a size array with a period of 1 µm was 30 min, while the production of an array with a period of 700 nm took 45 min. A tightly focused laser beam with a 1.1 µm spot and an objective featuring a numerical aperture of 0.5 were used. The Gaussian intensity distribution of the laser beam allowed us to achieve smaller diameters of the structures and periods of the arrays than the beam size. However, in this chapter, where we compare different shapes on various thicknesses of Au films, to ensure that the outcome is not affected by the beam overlapping, we investigate arrays with an equal 1 µm period.

As the laser pulse energy is increased, various morphological structures are produced. Based on their structural dimensions and characteristic shapes, the main four morphological states were distinguished: bumps, featuring a diameter greater than the apex height; cones, having similar lateral and height dimensions; jets, which have tips taller than their diameter; and crowns, featuring ablated holes in their centers. The bumps, cones, and jets are shown in Figure 1b. Using just enough energy to modify the Au layer, small roundly-shaped protrusions known as bumps are formed. The formation of these requires varying amounts of energy: 0.2 nJ, 0.4 nJ, 0.6 nJ, and 0.75 nJ for 25 nm, 50 nm, 75 nm, and 100 nm Au films, respectively. At higher pulse energies (0.3 nJ (25 nm), 0.75 nJ (50 nm), 1 nJ (75 nm), and 1.25 nJ (100 nm)), the protrusion height increases, and more pronounced cones are produced. At 0.4 nJ (25 nm), 0.95 nJ (50 nm), 1.25 nJ (75 nm), and 1.4 nJ (100 nm), we were able to produce jets with sharp, high, and narrow needle-like tips. Finally, when the energy levels exceed the point of jet splashing, the crowns and holes are fabricated.

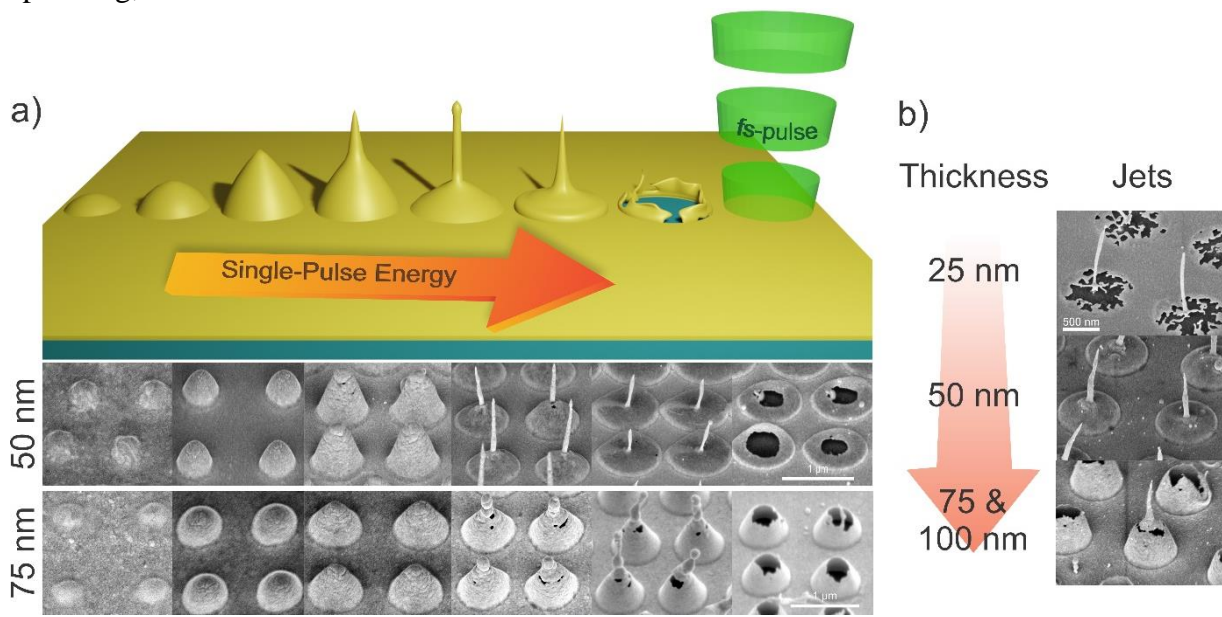

**Figure 2.** a) Illustration of nanostructure shape dependence on pulse energy and SEM images of nanostructures fabricated on 50 nm (upper row of SEM images) and 75 nm (lower row) thickness Au layer. b) SEM micrographs of jet shape dependence on Au thickness.

Figure 2a illustrates the energy-dependent morphological change with corresponding SEM images of the fabricated structures on 50 nm and 75 nm Au layers. The bumps and cones observed across all four different thicknesses of metal are quite similar. However, the formation of jets varies depending on film thickness (Figure 2b). For the thinner layers, the produced jets have sharp spikes that are narrow and high, standing either on the surface of the glass (25 nm) or on a cone or a disk-like platform that is made of a collapsed cone (50 nm). By contrast, for the thicker layers (75 nm and 100 nm), broader jets are formed at the top and in the center of the cone. Jets formed on the 25 nm Au layer manifestly have the thinnest spikes with a diameter-to-height ratio being 1:17, while on 50 nm, 75 nm, and 100 nm Au films the ratios are 1:13, 1:9, and 1:7, respectively. Differences in morphology may be attributed to the varying quantity of the metal in the spike. On the other hand, for the 25 nm and 50 nm Au layers, hydrothermal phenomena allow the molten part of the metal to flow towards the spike causing the collapse of the surrounding thin layer of the cone. When a thicker metal is used, there is more gold in the regions around the antenna and therefore the pedestal remains conical.

**2.2. Thermal effects induced in different Au thicknesses**

An analysis of the thermal effects following irradiation of the Au/SiO$_2$ two-layered complex at $E$=0.8 nJ for four values of distinctly different thicknesses ($d$=25 nm, 50 nm, 75 nm, 100 nm) has been conducted. The results of the induced spatial distribution lattice temperature are illustrated in Figure 3, along with the SEM images of the structures obtained at the same conditions. To evaluate the magnitude of the thermal effects, $T_L$ is illustrated in Figure 3 at three different time points (first column: at $6\tau_p$ when laser is switched off ($\tau_p$=300 fs), second column: when maximum temperature is attained, third column: at $t$=50 ps).

Theoretical predictions show that a decrease in the material thickness leads to higher lattice temperatures ($T_L$). This originates predominantly from the fact that smaller thicknesses result in a smaller depth in which the electrons can diffuse [56]. It is known that the electron system loses energy via two competing effects, diffusion and electron-phonon scattering. As the electron diffusion is inhibited for smaller $d$, hot electrons remain confined in a small volume and, thus, they lose energy, in principle, through electron-phonon scattering. Therefore, the electron temperature decrease is not as rapid as in the case of thicker materials. Thus, as the thickness decreases, the phonon subsystem will interact with an electron system, which is highly energetic and attains large temperatures. Interestingly, at $d$=25 nm, the maximum $T_L$ attained exceeds a critical temperature $T_L^c$ above which it is assumed that material is ablated (i.e. removed) (shown as a white region in the second and third column in Figure 3a). A thermal criterion was used to select $T_L^c$. More specifically, $T_L^c = 0.9 T_{cr}$ where $T_{cr}$=6250 K is the critical point for Au [57, 58] and is taken as the critical value of temperature above which ablation starts to occur. By contrast, for all thicknesses, a phase transition (i.e. melting) occurs in some parts of the material as the lattice temperature exceeds the melting point of Au (i.e. $T_{melt}$=1337 K). In that case, surface modification is via a mass displacement and not due to mass removal.

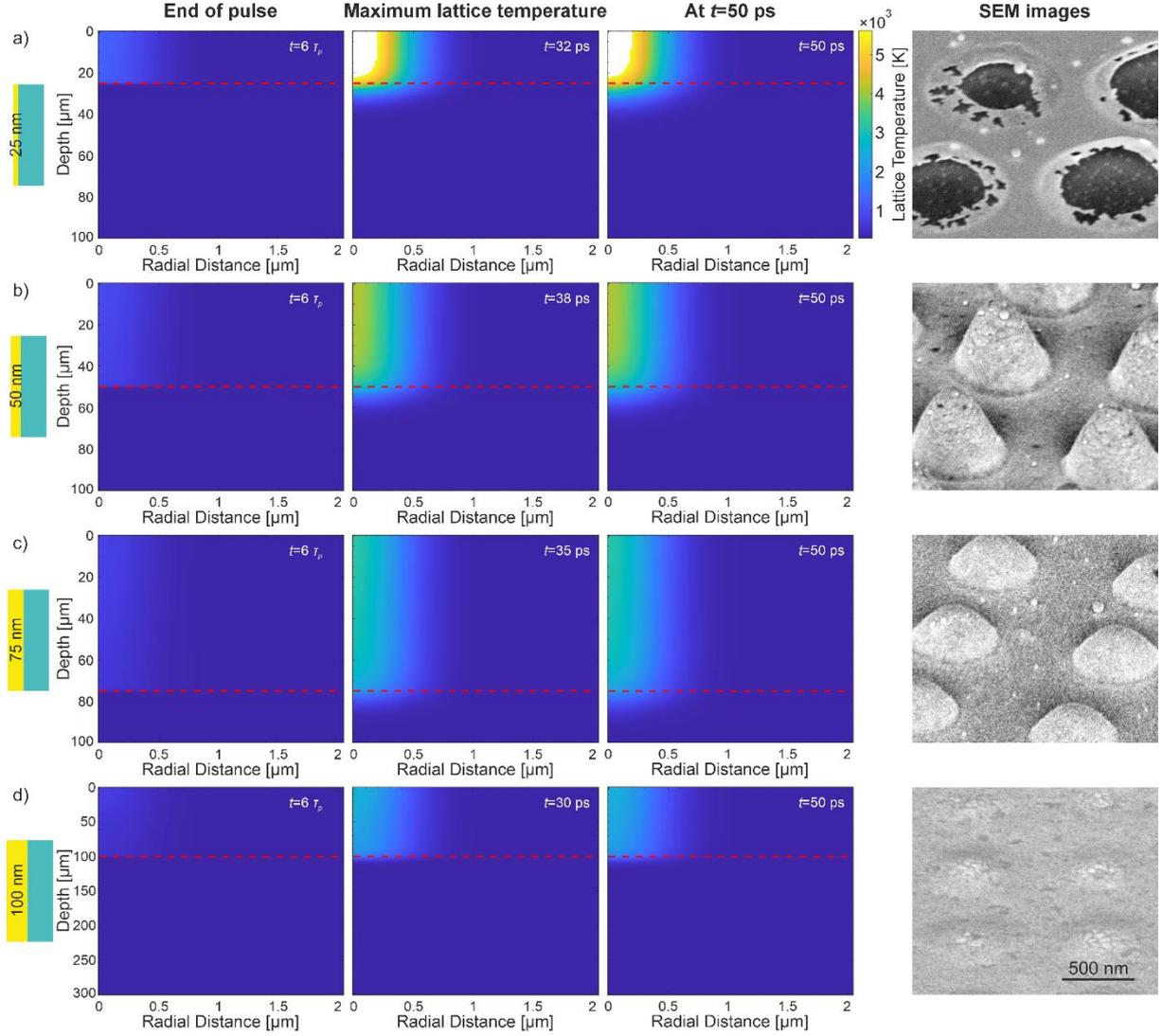

**Figure 3**. Lattice Temperature ($T_L$) profile in the propagation plane at three-time points: just after the end of the pulse ($6\tau_p$), when $T_L$ reaches the maximum value, and at $t=50$ ps. The first, second, and third column corresponds to these, respectively. The fourth column represents the SEM images of the practically obtained structures. The first, second, third, and fourth rows refer to the analysis of the $T_L$ for (a) $d=25$ nm, (b) 50 nm, (c) 75 nm and (d) 100 nm. The *red* dashed line stands for the interface between Au and $SiO_2$. *White* region in the first row (second and third column) indicates the ablated part of Au. All simulations were performed at $\lambda_L=515$ nm, $\tau_p=300$ fs, and $E=0.8$ nJ. The same colormap (min-max values) is used in all figures.

The simulated results correspond to the actual structures obtained using the exact parameters. SEM images of those are depicted in the fourth column of Figure 3. Ablated holes are obtained in the thinnest layer, while as the thickness increases, smaller protrusions are achieved due to phase transition occurring in a smaller part of the irradiated zone.

## 2.3. Tunable morphology of energy-dependent structures

To analyze the formation of the nanostructures, we utilized a focused Ga ion beam to create the cross-sections of the ones obtained in 75 nm thick Au. Corresponding SEM images can be observed in Figure 4. Three distinct morphological structures are all hollow inside, whereas the growing antenna appears to be solid. Comparing the internal height of these structures (cyan arrows in Figure 4), we observe an increase with growing bumps (in this case 322 nm) until the cones are reached (562 nm), when the growth rate for jets slows down (600 nm). At this point, the growth of the spike is rapidly increasing. To analyze the movement of melted material and the formation of specific structures, we measured the thickness of the film in both the lower periphery (green measurements) and upper parts closer to the center of the structure (yellow measurements). The bumps (Figure 4a) have a consistent thickness of approximately 40 nm over the entire cross-section. This suggests that the metal only detaches from the substrate and the movement of the melt has not yet begun. The wall thickness of the cones (Figure 4b) has decreased to 29 nm at the bottom (periphery) of the structure, while it has only slightly reduced at the top (37 nm). Additionally, a small antenna starts to grow at the top of the cone. This demonstrates that molten metal moves from the irradiated periphery to the center, resulting in the formation of a growing small solid tip at the top of the structure. As previously mentioned, such flow towards the center appears due to the temperature gradient and resultant induced velocity [40]. Here, the collision of the melt at the center has not yet fully started, thus the growing spike is minimal. In the jets (Figure 4c), the wall thickness is consistent at the bottom relative to the cones (27 nm), while the top layer becomes significantly thinner (20 nm), indicating that the impact of melt and further growth of the jet spike is mainly influenced by the movement of material from the areas surrounding the center. This significant thinning of the coating around the peak can explain the jet spike breakage at higher energies in thick coatings, as there is a lower film thickness in that area (Figure 4d).

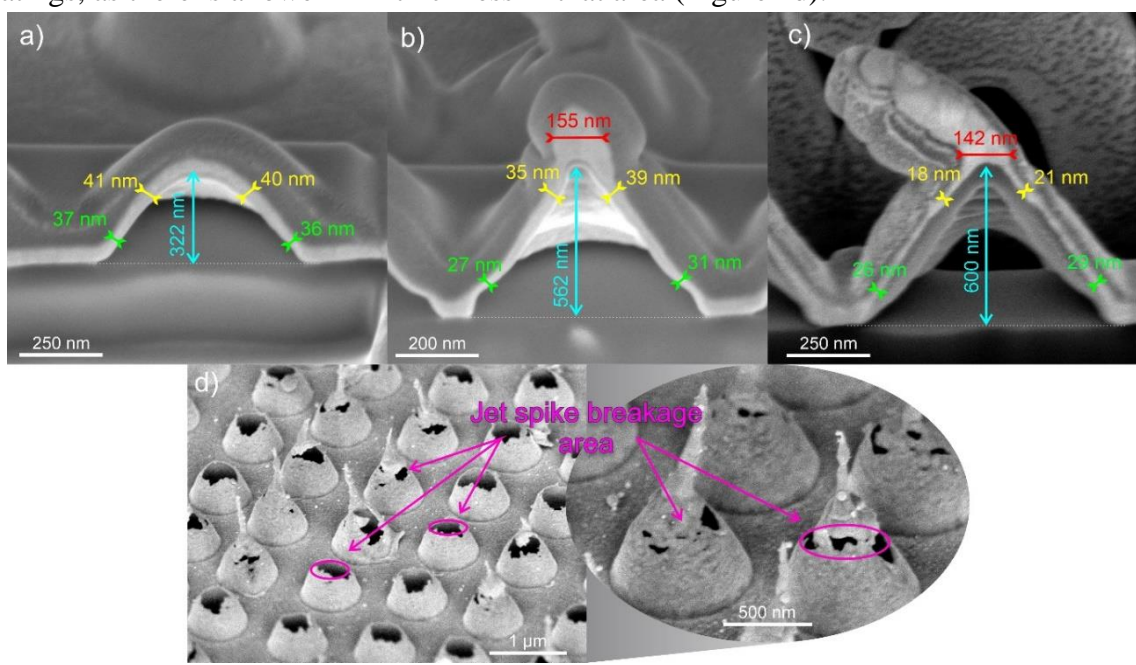

**Figure 4.** Gold bump (a), cone (b), and jet (c) cross-section with internal height and Au thickness measurements fabricated on 75 nm Au. The Au layer is in bright grey, while dark grey is platinum used for better contrast. Blue measurement shows the inside height of the structure, red – spike width, green – Au thickness in the periphery of the structure, yellow – thickness closer to the center of the structure. d) SEM images of broken jets on 100 nm Au with marked jet spike breakage area at the top of the structure, where the layer is the thinnest.

## 2.4. Thermal effects of different energy pulses

An analysis of the thermal effects following irradiation of the Au/SiO$_2$ two-layered complex for a 50 nm-thick Au layer placed on a SiO$_2$ substrate has also been conducted for five values of the laser energy ($E$=0.2 nJ, 0.5 nJ, 0.8 nJ, 1.1 nJ, 2 nJ) and results of the induced spatial distribution lattice temperature are illustrated in Figure 5. To evaluate the magnitude of the thermal effects, $T_L$ is illustrated in Figure 5 at three different time points (*first column*: at $6\tau_p$ when laser is switched off, *second column*: when maximum temperature is attained, *third column*: at $t$=50 ps).

Theoretical predictions show an anticipated increase of the lattice temperature of Au at increasing the laser energy. In particular, ablation occurs at $E$=2 nJ while only melting takes place at lower energies (at $E$=0.5 nJ). By contrast, for the lowest value of the energy ($E$=0.2 nJ), a phase transition and therefore a surface modification is not predicted. The simulation results are in agreement with the experimental observations illustrated in the SEM images assuming similar laser parameters.

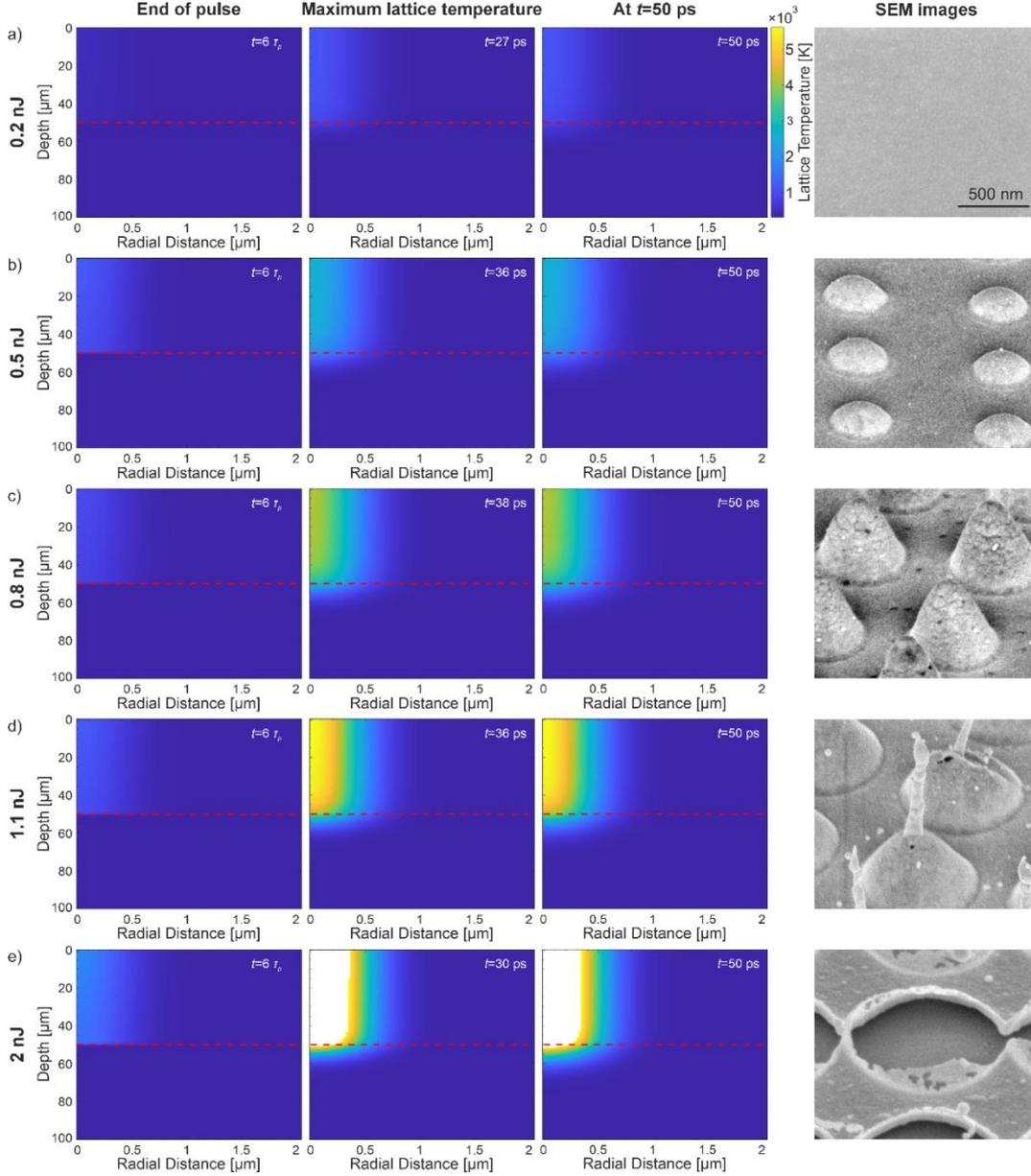

**Figure 5.** Lattice Temperature ($T_L$) profile in the propagation plane at three-time points: just after the end of the pulse ($6\tau_p$), when $T_L$ reaches the maximum value, at $t$=50 ps. The first, second, and third column corresponds to these, respectively. The fourth column represents the SEM images of the practically obtained structures. The first, second, third, fourth, and fifth rows refer to the analysis of the $T_L$ for (a) $E$=0.2 nJ, (b) 0.5 nJ, (c) 0.8 nJ, (d) 1.1 nJ, (e) 2 nJ. The *red* dashed line stands for the interface between Au and $SiO_2$ (assuming a 50 nm Au thick film). *White* region in the last row (second and third column) indicates the ablated part of Au. All simulations were performed at $\lambda_L$=515 nm and $\tau_p$=300 fs. The same colormap (min-max values) is used in all figures.

## 2.5. Grating period-induced tunability of structure shape

As previously mentioned, the Gaussian intensity distribution of the laser beam allowed us to achieve smaller diameters of the structures and array periods than the focused beam diameter. To fully understand and gain insight into the effect of the period on the formation energy and morphology, arrays with periods ranging from 0.5 µm to 1.3 µm were fabricated on all four different thicknesses.

After examining SEM images of each array, a map of the morphology dependence on energy and period was created and represented in Figure 6a. Firstly, we observe an energy-dependent change in morphology in the graph, with bumps at lower energies and jets at higher ones. Furthermore, the impact of the period is also demonstrated. For periods smaller than the focused beam diameter (1 µm, a decrease in distance at a constant energy results in the fabrication of the higher-energy requiring structures. Cones are formed by decreasing the bump array period, whereas by reducing the cones' grating inter-structural distance, jets can be produced. The impact of the period on shape is lower than that of pulse energy, thus changing only the distance with constant energy will not lead to the fabrication of all morphological states. Additionally, Figure 6b illustrates the comparison of different period gratings fabricated on various thicknesses of Au using the same pulse energy. The alterations of both distance and thickness lead to an evident change in the structure shape. The full energy-period morphology change maps for 25 nm, 50 nm, 75 nm, and 100 nm can be found in Figure S1, Figure S2, Figure S3, and Figure S4, respectively.

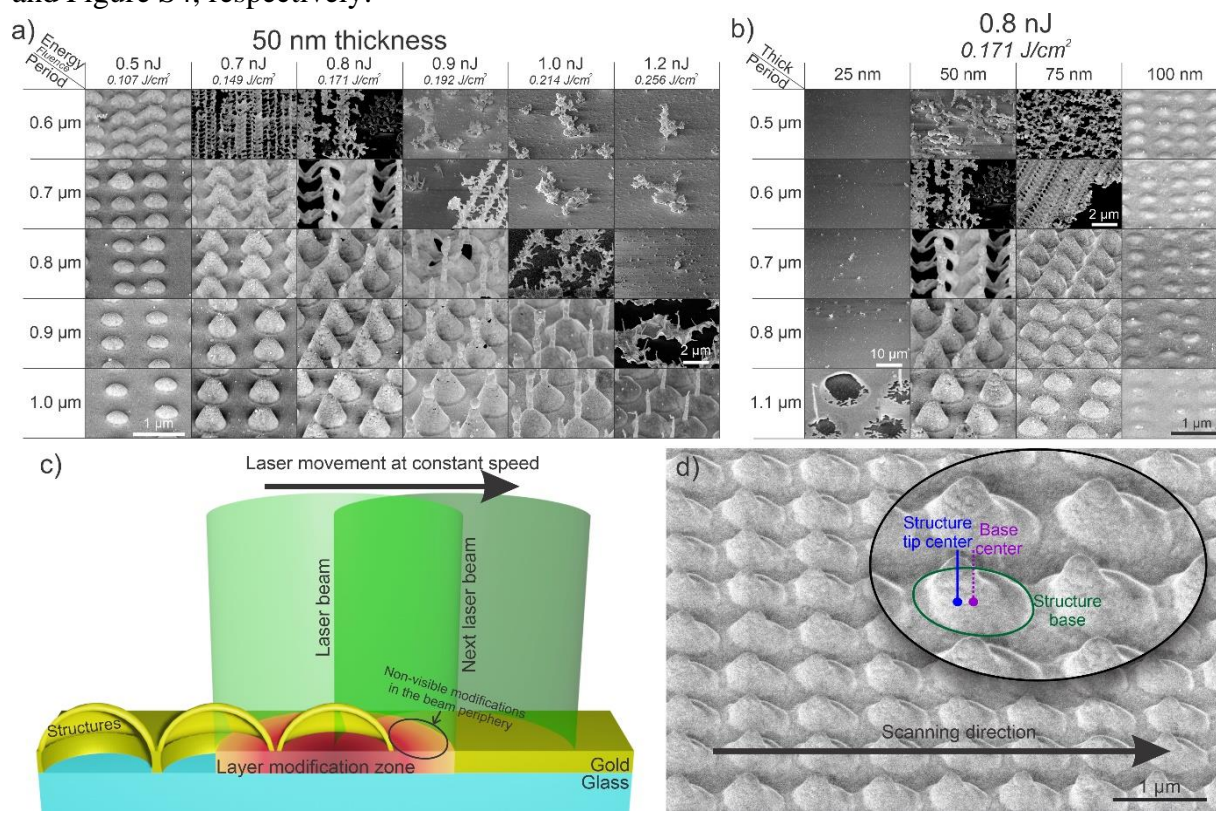

**Figure 6.** a) Morphology dependence on period and energy/fluence map for 50 nm thickness Au. The period is in rows, while energy is in columns. b) Morphology dependence on period and Au thickness map for constant energy 0.8 nJ (0.171 J/cm$^2$) pulses. c) Illustration of the overlapping effect of the laser beam and the layer modification zone in arrays with periods smaller than the focused laser beam diameter. d) The shift of the structure's tip to the previously modified area in the opposite direction of scanning.

Our results demonstrate that by utilizing a sub-diameter period the energy required to fabricate specific shape structures can be lowered. This reduction occurs because a smaller period causes higher overlapping of the irradiated area, leading to fabrication within a previously modified zone. As the laser beam hits the metal surface, the affected layer gains mechanical stresses throughout the whole irradiated region. This effect cannot be attributed to the heat accumulation [59, 60], as the time between adjacent pulses is quite long (i.e. 7 mm/s scanning speed with 10 kHz pulse repetition rate, leading to 100 µs between successive pulses). Induced stresses in the peripherical parts of the irradiated area are not high enough to visibly lift the film. Thus, the surface modification in the central area leads to the formation of protrusions, while in the periphery the non-visible slight delamination or intrinsic modifications in the metal are obtained. They only affect the fabrication of the next structure formed with the subsequent overlapping beam pulse, leading to the reduction of the required fluence by reducing the period. Beam overlapping effect is represented in Figure 6c. Meanwhile, if the grating period exceeds the diameter of the focused beam, the shape and size of the structures remain constant and are not affected by the distance (can be observed in Figure S1-S4 at periods larger than 1.1 µm). In addition, due to the higher overlapping, the shape appears inhomogeneous and non-uniform in the scanning axis. The tip of the structure is formed with a shift towards the previously modified area (already formed structure) relative to its base center (Figure 6d). Such non-centrality is only observable at higher fluences, when part of the affected layer is melted.

## 2.6. Structure formation thresholds

In order to determine the formation parameters more accurately, the diameters of all the structures were measured and the dependence of squared diameter on the natural logarithm of formation energy for four distinct Au layer thicknesses was plotted in Figure 7a. The differently colored lines illustrate this dependence for different grating periods. By linear approximation of the graph data, the practical layer modification threshold energy and the characteristic energy deposition diameter (area in which laser energy is effectively absorbed for the formation of single-pulse induced nanostructures) can be calculated. The latter diameter is obtained from the slope of the approximated line, while its intersection with the energy axis indicates the minimum energy required for layer modification.

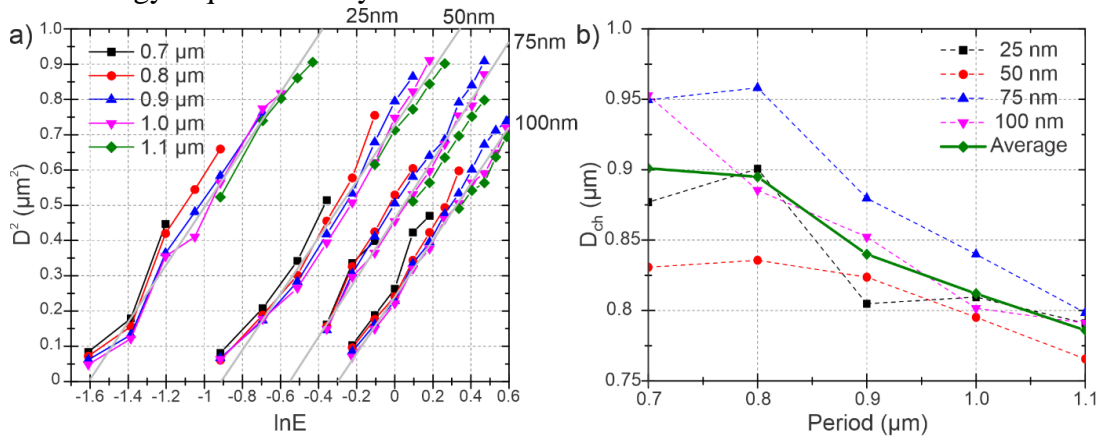

**Figure 7.** a) Squared diameter dependence on the natural logarithm of pulse energy for 25 nm, 50 nm, 75 nm, and 100 nm Au layers. The different colors depict different grating periods. The linearly approximated 1 µm period array data is represented by a gray line. b) Energy deposition diameter dependence on the grating period. The colored dashed lines depict different layer thicknesses. Solid line represents the averaged values of all different thicknesses for each period.

The data dependence for all thicknesses and periods is nearly linear. Comparing the diameters of the structures formed with the same energies but different periods, a tendency of a slight increase in size by reducing the period can be observed, which is attributed to the overlapping effect described above. The characteristic energy deposition diameter dependence on the grating period for each thickness is represented in Figure 7b as dashed lines, with the averaged value for each perdiod as well (green solid line). Here, the general trend of the energy deposition diameter reduction with period increase is observed – from averaged (green solid line in Figure 7b) $D_{ch}$ = 0.9 µm at highly-overlapping 0.7 µm and 0.8 µm arrays to $D_{ch}$ = 0.79 µm at non-overlapping 1.1 µm arrays. This change confirms the overlapping effect, indicating that a larger modified zone (structure with a larger diameter) at smaller periods corresponds to a greater-diameter zone at which energy is effectively absorbed for the structure formation. Also, it is worth mentioning that this characteristic energy deposition diameter ($D_{ch}$ = 0.79 µm) is smaller than the theoretically calculated Gaussian beam spot size ($D_{opt} = \frac{4 \cdot M^2 \cdot \lambda \cdot f}{d \cdot \pi}$), which in this case is 1.1 µm. Such a discrepancy could be attributed to the idea that the effective energy absorption for structure formation occurs in a region of smaller radius than the Gaussian beam size (in its center) where the threshold is crossed.

To determine the modification energy while excluding the overlapping effect, the data from the arrays with a period of 1.1 µm was linearly approximated. This approximation is marked in Figure 7a as a gray line. The computed layer modification energy and fluence for each thickness are represented in Table 1.

**Table 1.** Different thickness thin Au layer modification threshold.

| **Thickness, nm** | **25** | **50** | **75** | **100** |
|---|---|---|---|---|
| **lnE** | -1.61 | -0.91 | -0.545 | -0.295 |
| **$E_{th}$, nJ** | 0.2 | 0.4 | 0.58 | 0.74 |
| **$F_{th}$, J/cm$^2$** | 0.043 | 0.085 | 0.124 | 0.158 |

In the thinnest 25 nm coating, the line intersects the abscissa axis at ln$E$=-1.61, corresponding to energy $E_{th}$=0.2 nJ. The modification threshold for thicker 50 nm, 75 nm, and 100 nm coatings stands at 0.4 nJ, 0.58 nJ, and 0.74 nJ respectively. By comparing the threshold energies at 50 nm and 25 nm, the change of it is 2 times, while in thicker 100 nm and 75 nm this change is only 1.28 times. These results show that in thinner coatings, the modification energy changes more rapidly based on the thickness and the coatings are more sensitive to energy fluctuations.

The fluences for morphological state transitions were determined depending on the period by utilizing the data and maps from Figures S1-S4. The transitional fluences are graphically represented in Figure 8. The green line indicates the threshold for the bump formation and layer modification, which was obtained from the linear approximation in Figure 7a. The red line represents the fluence at which the morphological phase transition occurs from bumps to cones. The blue line shows the fluence at which cones transition to jets. The pink line indicates the critical fluence at which the destruction of the structures occurs, along with the formation of crowns.

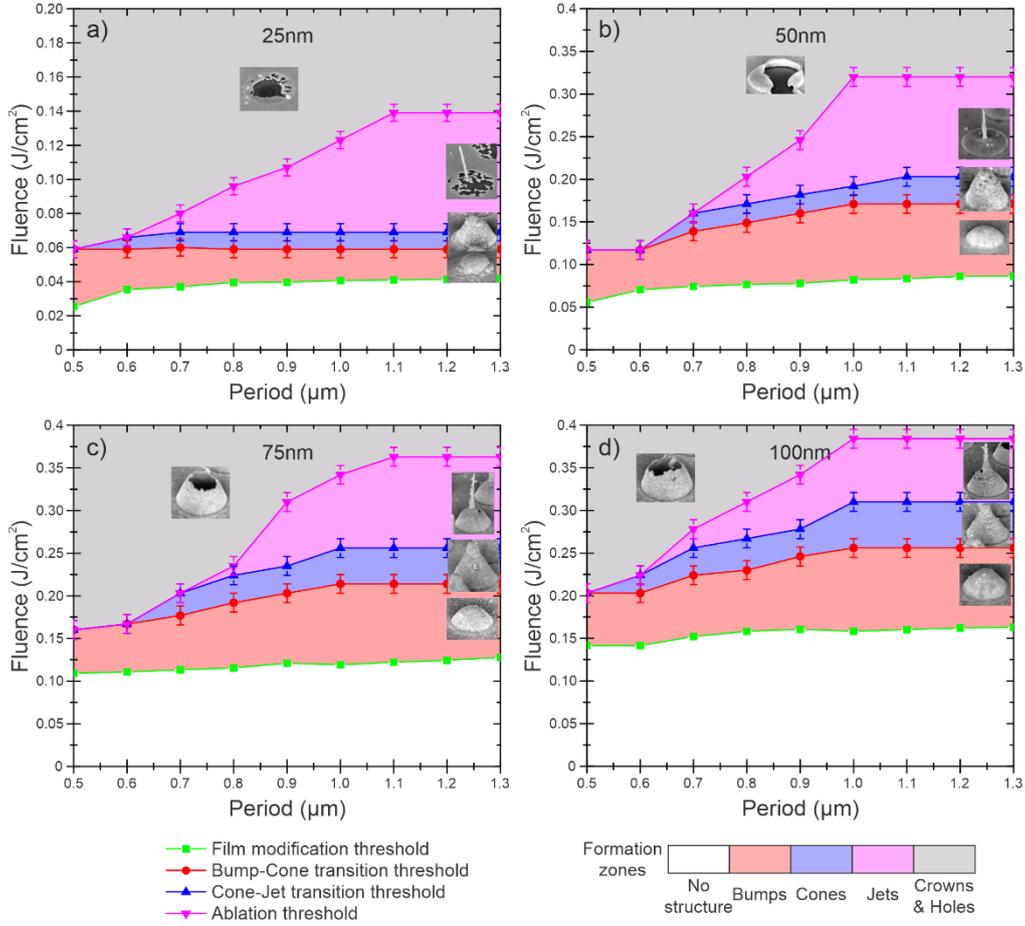

**Figure 8.** Morphological state transition energies: layer modification and bump formation (green line), bump transition to cone (red line), cone transition to the jet (blue line), and layer ablation (pink line) energies in 25 nm (a), 50 nm (b), 75 nm (c) and 100 nm (d) layer.

Additionally, the graphs demonstrate a period-influenced overlapping effect, manifested by changes in ablation and shape transition fluences in arrays with a period smaller than the beam size. At higher periods, the fluences remain constant; however, as the period decreases, these decrease as well. The range of fluences where the bumps are fabricated (red zone) is quite broad and can be achieved in all gratings. On the other hand, cones and jets are not produced at extremely small periods (0.5 – 0.6 µm) and only appear in arrays with a pitch larger than 0.6 µm due to excessive overlapping. In addition, the formation energy range of the cones is the narrowest compared to that of the bump-like and jet-like structures. This is because only structures with similar heights and diameters are considered as the latter morphological state.

**2.7. Dependence of plasmonic properties on morphology and layer thickness**

The fabricated 3x3 mm² periodic gratings generate hybrid lattice plasmon resonance (HLPR), which can be observed as a drop in the reflectance intensity. The reflectance spectra of the arrays were measured using Essentoptics Photon RT UV-VIS-NIR spectrophotometer in the range of 400-1600 nm wavelength with the step of 4 nm. A circular-shaped, 2 mm diameter incident beam was used, and the spectra were measured for both *s*- and *p*-polarized light, with a constant 8° angle of incidence. Due to the non-uniform formation of arrays in perpendicular axes, the highest-quality results were achieved when the electric field was perpendicular to the scanning line. Therefore, for different polarizations, the sample was rotated along an azimuthal axis. Figure 9a-b displays the reflection spectra of four types of nanostructures – small bumps (height of up to 100 nm), normal bumps, cones, and jets, on a 100 nm thick Au layer with a 1 µm period. Here, mainly the 3 resonant peaks are investigated that correspond to the perpendicular grating of 1 µm: in *p*-polarization, two peaks corresponding to $m=+1$ (875 nm) and $m=-1$ (1147 nm) diffraction order, and in *s*-polarization one peak corresponding to $m=\pm1$ (1000 nm) diffraction order. Other observed peaks are either for higher diffraction order or diagonal grating plane, but these are not investigated. The spectra of the matrix of small bumps is barely distinguishable from an ordinary reflection of plain Au in both *s*- and *p*-polarization and only shallow and narrow HLPR dips are visible. For higher bumps, the narrow dip corresponding to $m=-1$ diffraction order becomes intense, with a slight shift to higher wavelengths (1204 nm) for *p*-polarized light in comparison with smaller structures, while the wavelength remains the same for $m=+1$ diffraction order (876 nm) and *s*-polarization. The resonant wavelength for this type of grating can be calculated theoretically using the diffractive grating formula [51]. For higher cones, this model is not applicable as the $m=-1$ resonant wavelength shifts towards longer wavelengths at *p*-polarization (1288 nm) and is itself wider and shallower due to the excitation of several hot spots (significant enhancement of local electric field) at the base and tip of larger structures. Arrays of sharp jets have a much wider HLPR with either a significant redshift or a strong decrease in the total reflection in the whole measured range, making induced resonance unobservable. This same tendency is observed across all 4 different thicknesses.

The main aspect is that the resonance shift is only observed for one peak – at longer wavelengths in *p*-polarization. Such shift of the resonant wavelength with increasing height of the structure was previously explained by the increased effective period [61], which covers not only the distance between the structures but also the surface length of the protrusion (Figure 9c). Bumps feature a smoother and shorter top surface, whereas cones and jets possess longer surface distances. Consequently, the larger effective period contributes to the resonance shift. However, the shift with changing height might suggest that this spectral feature can be associated with the out-of-plane (*z*-axis) electric component of the incident light. Since the *p*-polarized light incident at an oblique angle covers the vertical component of the electric field, it excites height-dependent out-of-plane plasmonic modes [62-64]. The out-of-plane mode is determined to suffer from high Ohmic losses at large heights of the structures, which might explain the results of the jet spectrum. The perpendicular component excites in-plane modes that mainly depend on the array period and structure diameter.

The broadening of the peak can be attributed to the non-uniformity of the nanostructures in the array. As previously discussed, the change in height leads to a red-shift. The bump structures are quite uniform and the collective resonance occurs at a similar wavelength. With the application of higher fluence and the fabrication of cones, the resulting structure height is more susceptible to fluctuations in pulse energy. A larger distribution of heights results in a distribution of resonant wavelengths and a broadening of the total resonant peak. In jet arrays, each structure is slightly different (jet length, inclination), thus wide variation in height can lead to an extreme broadening of the resonance. Additionally, as it was observed in SEM images (Figure 4d), some structures in jet arrays were broken, which might reduce the total intensity of the reflection.

For theoretical estimation, we used COMSOL Multiphysics modeling to calculate the reflection spectra and near-field profiles of the structures at the main 3 investigated dips (880 nm, 996 nm, and red-shifting 1150-1300 nm). The sizes of the structures for simulations were determined from SEM images. Figure 9d-e shows the calculated reflection spectra for the bump and cone arrays. Meanwhile, the reflection spectrum of the jets is not included because the theoretical calculations do not match the practical results; in fabricated arrays, each structure varies slightly in terms of jet length or inclination, and some structures may even be broken. Therefore, the real spectrum is completely different from the theoretical one, and it is difficult to model such an array. The calculated results are in quite good agreement with the practical ones and the red-shift of the longer-wavelength resonance is also observable. The increase in structure height (either bump or cone) leads to a further change in the wavelength. Near-field profiles in the insets of Figure 9d-e additionally give an insight into the nature of these plasmonic modes. Red-shifting resonance at longer wavelengths tends to enhance the local field at the top of the structure suggesting the out-of-plane nature of this peak. Other investigated peaks exhibit enhancement at the sides of the structure coinciding with the in-plane excitations.

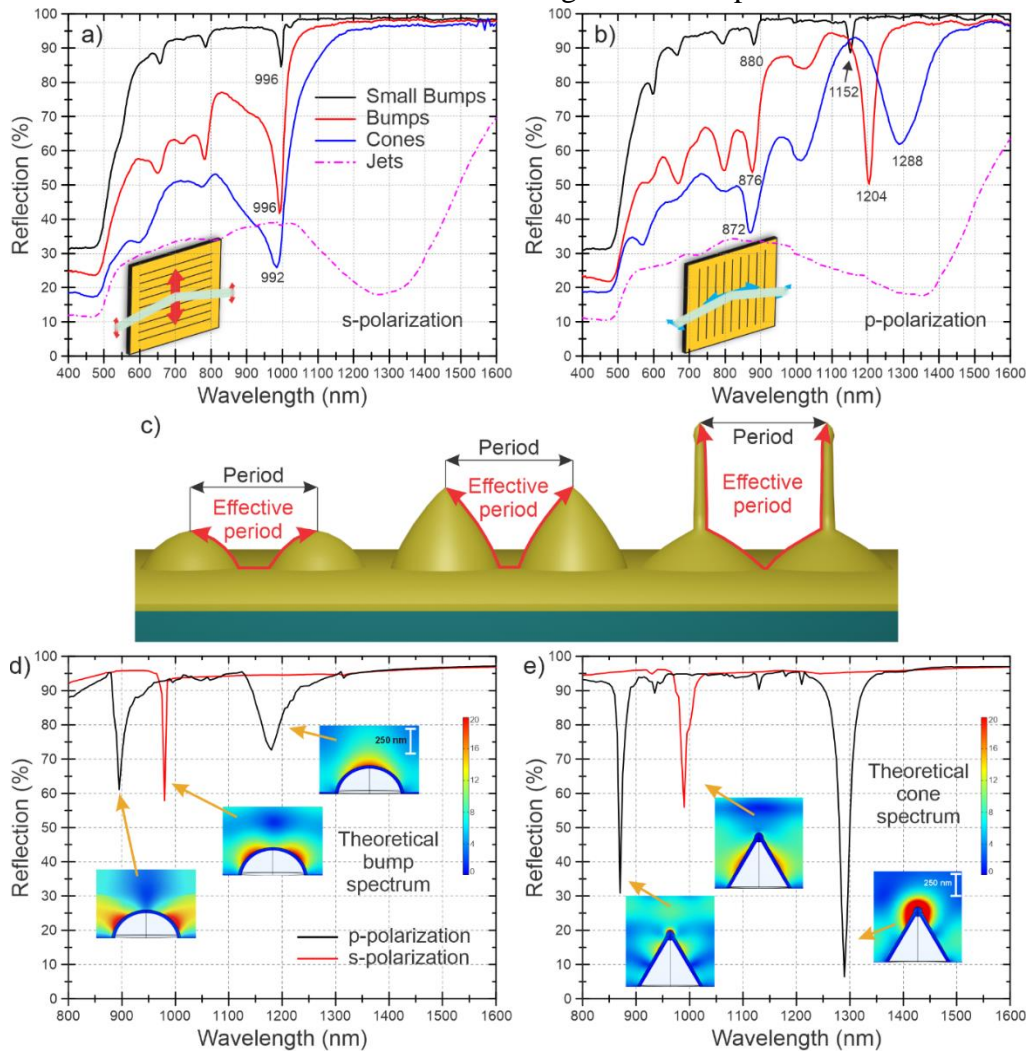

**Figure 9.** The reflectance spectra of different morphological structures: small bumps (black line), bumps (red), cones (blue), and jets (purple) arrays of 1 µm period on 100 nm thickness Au for *s*- (a) and *p*-polarization (b) at 8° light incidence angle. The insets show the azimuthal rotation of the sample. c) Grating period and effective period comparison for different shape structures. Numerically calculated reflection spectra for 1 µm period bump (d) and cone €  arrays on 100 nm gold. The insets show the near-field profile at the main 3 investigated resonant peaks.

The main quality parameters of the resonance – width and depth were calculated for matrices of various thicknesses. The depth was measured from the upper-left corner of the most prominent peak and its full width at half minimum (FWHM) was determined. Quality measurements included the calculation of quality (Q)-factor ($Q = \lambda_{res}/\Delta\lambda$), where $\lambda_{res}$ is resonant wavelength and $\Delta\lambda$ is FWHM, and the modified quality (MQ)-factor ($MQ = \frac{\lambda_{res} \cdot h}{\Delta\lambda \cdot 100\%}$), which includes the depth $h$ of the resonant peak. The results of the 1 µm gratings with various structure shapes and Au thicknesses for both *s*- and *p*-polarizations at 8° incident light angles are represented in Figure 10. The factors of *s*-polarization results were measured for small bumps, bumps, cones, and jets, while for *p*-polarized light the jet resonances were not determined due to high red-shift. By increasing the formation energy and changing the morphology, the Q-factor from *Q*=100 at small bump gratings decreases by tenfold in cone and jet arrays. Nonetheless, the MQ-factor for *s*-polarization is quite similar in small bump and normal bump arrays, whereas for *p*-polarization the normal bumps exhibit higher quality. The misfit between factors shows that although the Q-factor is higher in smaller structures due to the narrow FWHM, the more practical HLPR is observed in higher bumps arrays. This is due to the higher intensity resonance and larger intensity-width ratio. In addition, while there is no significant difference in distinct thicknesses, the 25 nm Au film yields the lowest quality factors, as the structures are small and thin, and the layer itself possesses the lowest reflectivity.

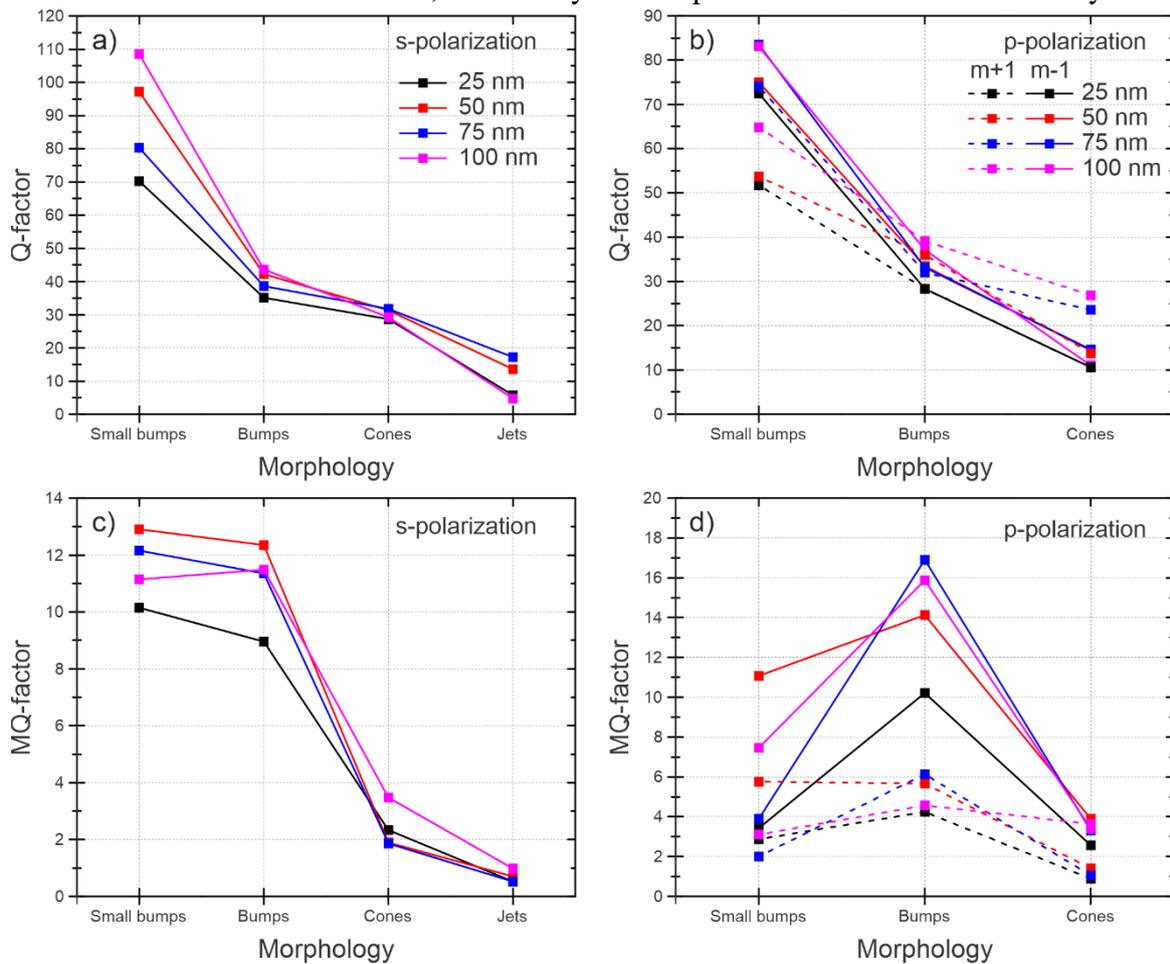

**Figure 10.** The 1 µm period grating resonant peak Q-factor (a, b) and MQ-factor (c, d) dependence on the shape of the structure and Au thickness for *s*- (a, c) and *p*-polarization (b, d) at an 8-degree light incidence angle.

Furthermore, we investigated the impact of reducing the period below the beam diameter on the quality of the resonance. Figure 11 displays the reflectance spectra of 0.7 µm and 1 µm bump gratings on 50 nm Au film. As the period was reduced, the resonance peak's depth and quality increased despite the observed resonant wavelength shift. The Q-factor increased from $Q$=33.5 at 1 µm grating to $Q$=43.3 at 0.7 µm grating in the *p*-polarization measurement, while the MQ-factor changed from $MQ$=14.1 to $MQ$=23. Meanwhile, in the case of *s*-polarization, the MQ-factor increased slightly and the Q-factor decreased due to a reduction in the resonant wavelength and a consequent under-reduction of FWHM. When the period is reduced, the interaction among nanostructures intensifies and the HLPR becomes stronger. This is because there are more interacting structures in the same irradiated area.

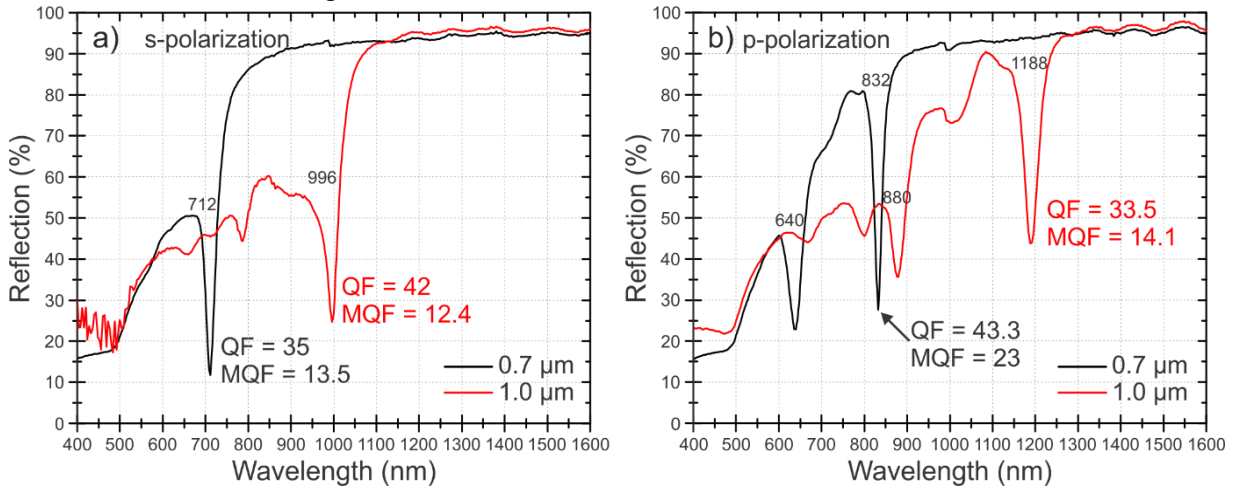

**Figure 11.** The 0.7 µm (black line) and 1 µm (red line) period bump grating reflection spectra for *s*- (a) and *p*-polarization (b) at 8-degree light incidence angle on 50 nm Au film.

## 3. Conclusions

It is shown that the morphology of single ultrashort laser pulses-induced nanostructures can be controlled via an appropriate selection of the pulse energy, thin metal film thickness, and array period. Thin films provide the formation of sharp and narrow jet-like structures, while the thickening of the material changes the formation principle, and shorter but wider structures are obtained. The melt flow during the formation of different morphology structures was also investigated and it was shown that at small energies the molten material moves from the sides of the structures to the central part, while at higher the movement from the central part to the tip occurs. The simulations prove that during all morphological stages, the formation of the material undergoes a phase transformation, and only reaches the ablation threshold when the crowns are formed. Further manipulation in shape can be achieved by reducing the grating period below the size of the beam diameter, as an overlapping effect takes place, which has been studied in detail. Finally, the morphology change offers new opportunities to control plasmon resonance, as the resonant wavelength and its quality change with structure shape, film thickness, and grating distance, thus providing an additional degree of freedom for the tunability of the plasmonic properties.

## 4. Experimental Section

*Sample preparation*: A magnetron sputter coater (Quorum Q150T) was used for the deposition of the thin Au films (25 nm, 50 nm, 75 nm, 100 nm) on soda lime glass. The film thickness was controlled by changing the deposition time from 94 s to 376 s with a constant sputtering rate of ~0.27 nm s$^{-1}$.

*Laser processing*: The periodic structures were fabricated using single pulses generated by Yb:KGW based femtosecond laser (Pharos, Light Conversion Ltd.). The second harmonic (515 nm) of the laser was used and the DLW technique was employed, tightly focusing the laser beam into a 1.1 µm spot using an objective featuring a numerical aperture of 0.5. The size of the fabricated arrays for SEM imaging was 100×100 µm$^2$ and for reflectance measurements was 3×3 mm$^2$. The range of investigated periods was 0.5-1.3 µm.

*Characterization*: The structures in the gratings were characterized using a scanning electron microscope (Helios NanoLab650). The diameters were measured from the images tilted at a 57° angle, using the image processing tool ImageJ. The morphology was distinguished according to the height-diameter ratio: structures with diameters larger than the height were considered bumps, the ones having similar dimensions (ratio close to 1) were cones, while those with high and narrow spikes were jets. The cross-sectional cuts of the structures were performed using a focused Ga ions beam, followed by a Pt layer deposition on the top using FIB-SEM Helios NanoLab650. The plasmonic properties of the gratings were characterized using a spectrophotometer (Photon RT, Essentoptics). The reflectance spectra for linearly polarized light were measured in the wavelength range of 400-1600 nm using 8° angle of incidence.

*Theoretical model*: To describe the effects on the Au/SiO$_2$ two-layered complex following irradiation with fs pulses, a theoretical framework is employed to explore the excitation and thermal response of the structure. The simulation algorithm is based on the use of a Two Temperature Model (TTM) that represents the standard approach to evaluate the dynamics of electron excitation and relaxation processes in solids [65]. In this work, the response of the structure with laser pulses of wavelength $\lambda_L$=515 nm and pulse duration equal to $\tau_p$=300 fs is simulated assuming a Gaussian beam of radius ($r_0$=0.68 µm). The intensity profile of the laser beam is provided by the following expression:

$$S = \frac{(1-R-T)\sqrt{4\log(2)}F}{\sqrt{\pi}\tau_p(\alpha^{-1}+L_b)} \frac{1}{(1-\exp(-d/(\alpha^{-1}+L_b)))} \exp\left(-4\log(2)\left(\frac{t-3t_p}{t_p}\right)^2\right)\exp\left(-2\left(\frac{r}{r_0}\right)^2\right)$$

where $R$ and $T$ stand for the reflectivity and transmissivity, respectively, $L_b$ corresponds to the ballistic length for Au (~100 nm [66]), $\alpha$ is the absorption coefficient for Au, and $F$ is the peak fluence of the laser beam ($F = \frac{2E}{\pi r_0^2}$) where $E$ stands for the laser energy. On the other hand, $t$ is the time, assuming the laser beam is switched on at $t$=0 and it is switched off at $t$=6$\tau_p$ and $r$ is the radial distance from the spot center. Due to the small thickness of the superstrate (Au), the optical parameters of the complex (Au/SiO$_2$) are expected to be influenced by the optical characteristics of both Au and SiO$_2$. Thus, a multireflection theory is employed to calculate these parameters which determine the excitation level in Au. Thermal effects occur in both materials; however, the transmitted energy is not sufficient to produce excitation to the substrate. A more detailed investigation of the excitation and thermal effects in the Au/SiO$_2$ is presented in Ref. [56].

*Numerical Simulation:* The reflection spectra and near-field profile of the plasmonic nanostructures were calculated using the finite element method implemented in COMSOL Multiphysics solver. The wavelength ranging from 800 to 1600 nm (the zone of interest for the main 3 investigated peaks) with a step of 5 nm was used to irradiate the structures at an 8° angle of incidence. The structure was described in a rectangular domain with periodic boundary conditions (Floquet periodicity) in both axes (1 µm period). To avoid boundary reflections, two perfectly matched layers were applied to the top and bottom of the domain. The structures were described as hollow ellipsoids or truncated cones with very short (for cones) or long (for jets) solid elliptical tips on a 100 nm gold layer. The dimensions of the structures were taken from SEM images. Johnson and Christy's data [67] was used for the dielectric function of gold. The thin film and structure were located on a soda-lime glass substrate with a determined thickness of 1 µm. Air domain was used above the structure with its thickness being double the used wavelength.

**Acknowledgements**

E.S and K.V. has received funding from the Research Council of Lithuania (LMT, Lithuania) under the project No. S-MIP-23-32 and from the Nanoscience Foundries and Fine Analysis – Europe (NFFA-Europe) under the project ID 369.

# Supplemental material

**Formation of Highly Tunable Periodic Plasmonic Structures on Gold Films Using Direct Laser Writing**

*Kernius Vilkevičius\*, George D. Tsibidis, Algirdas Selskis, Emmanuel Stratakis, and Evaldas Stankevičius*

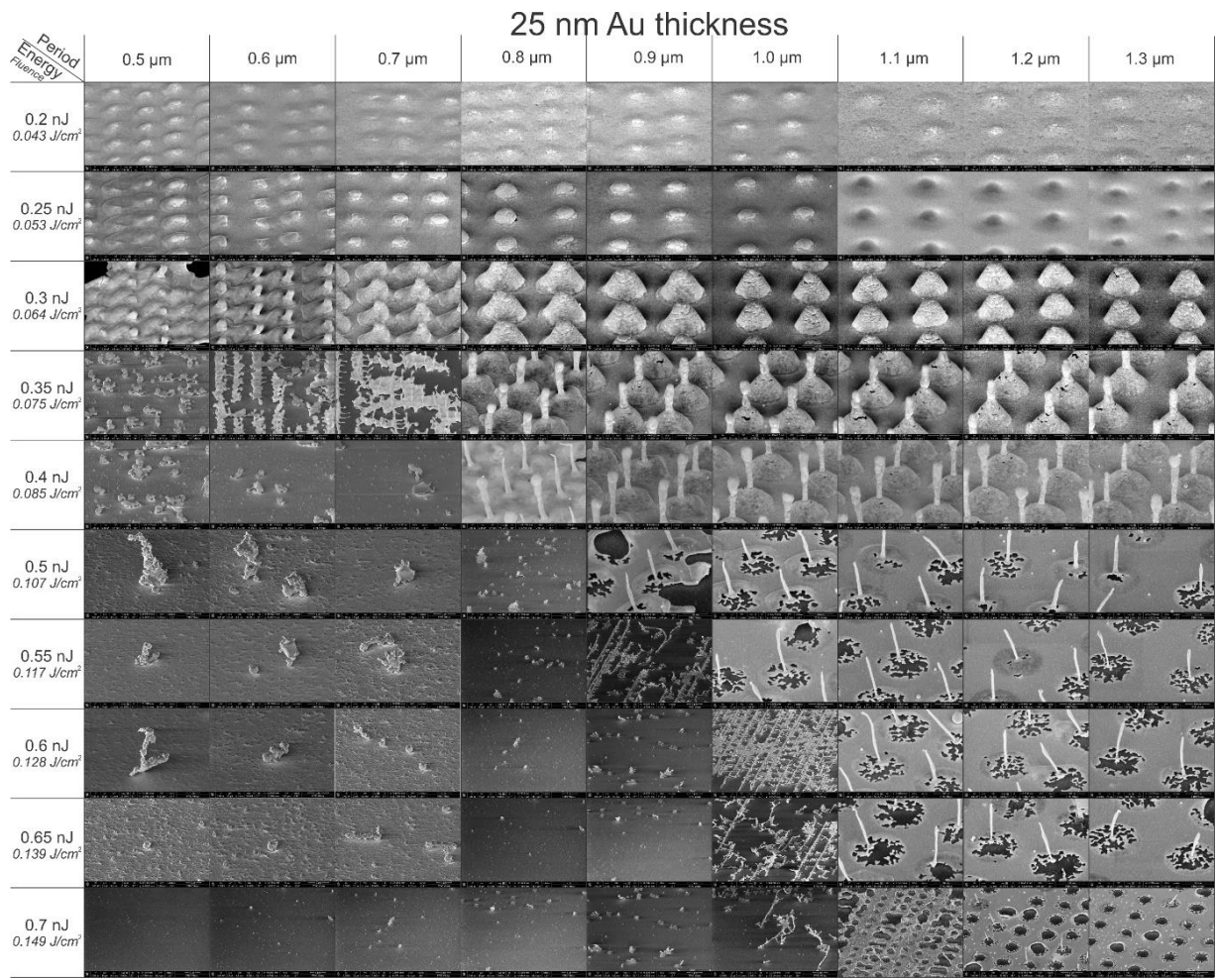

**S1.** Morphology dependence on period and energy/fluence map for 25 nm thickness Au.

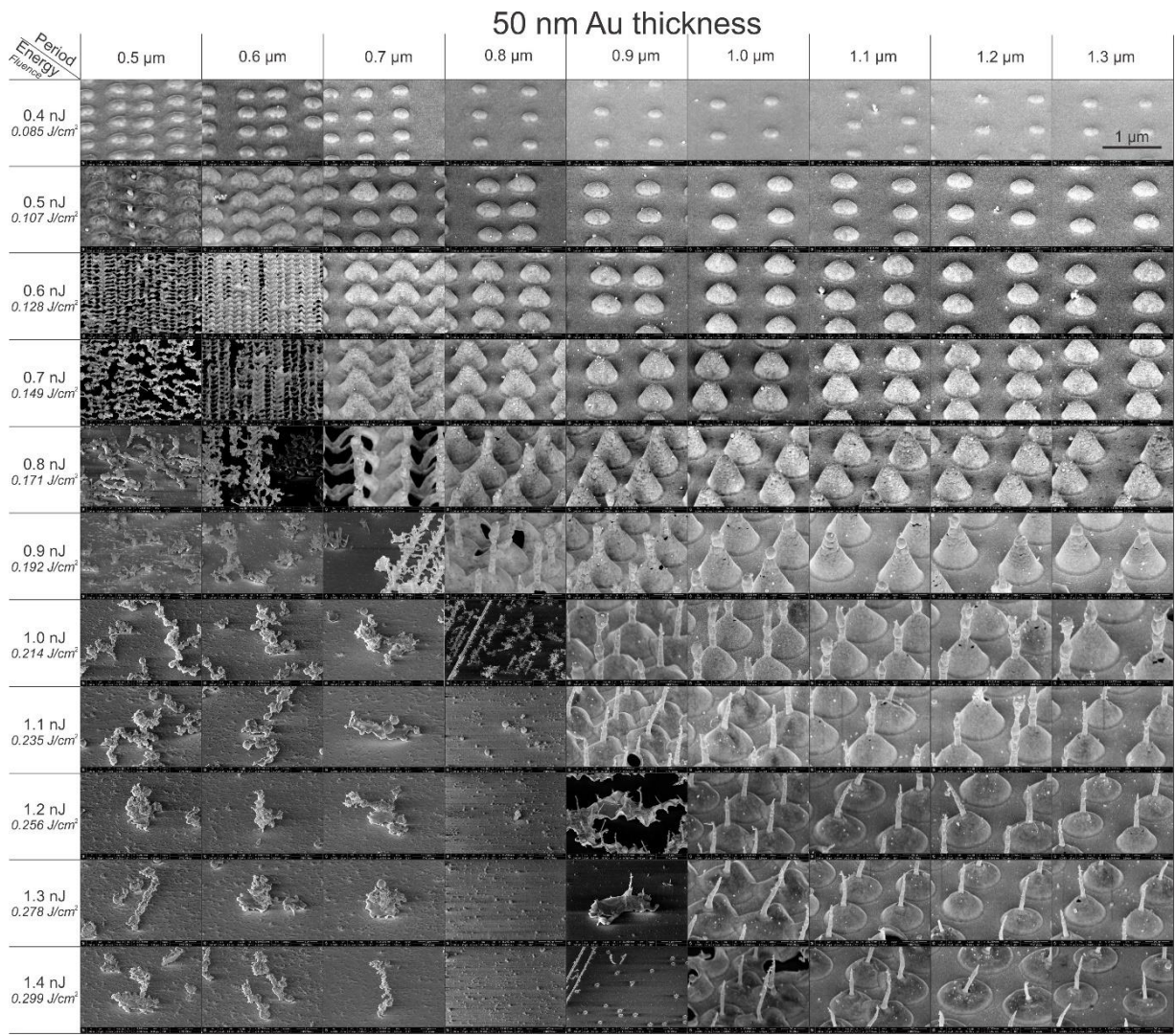

**S2.** Morphology dependence on period and energy/fluence map for 50 nm thickness Au.

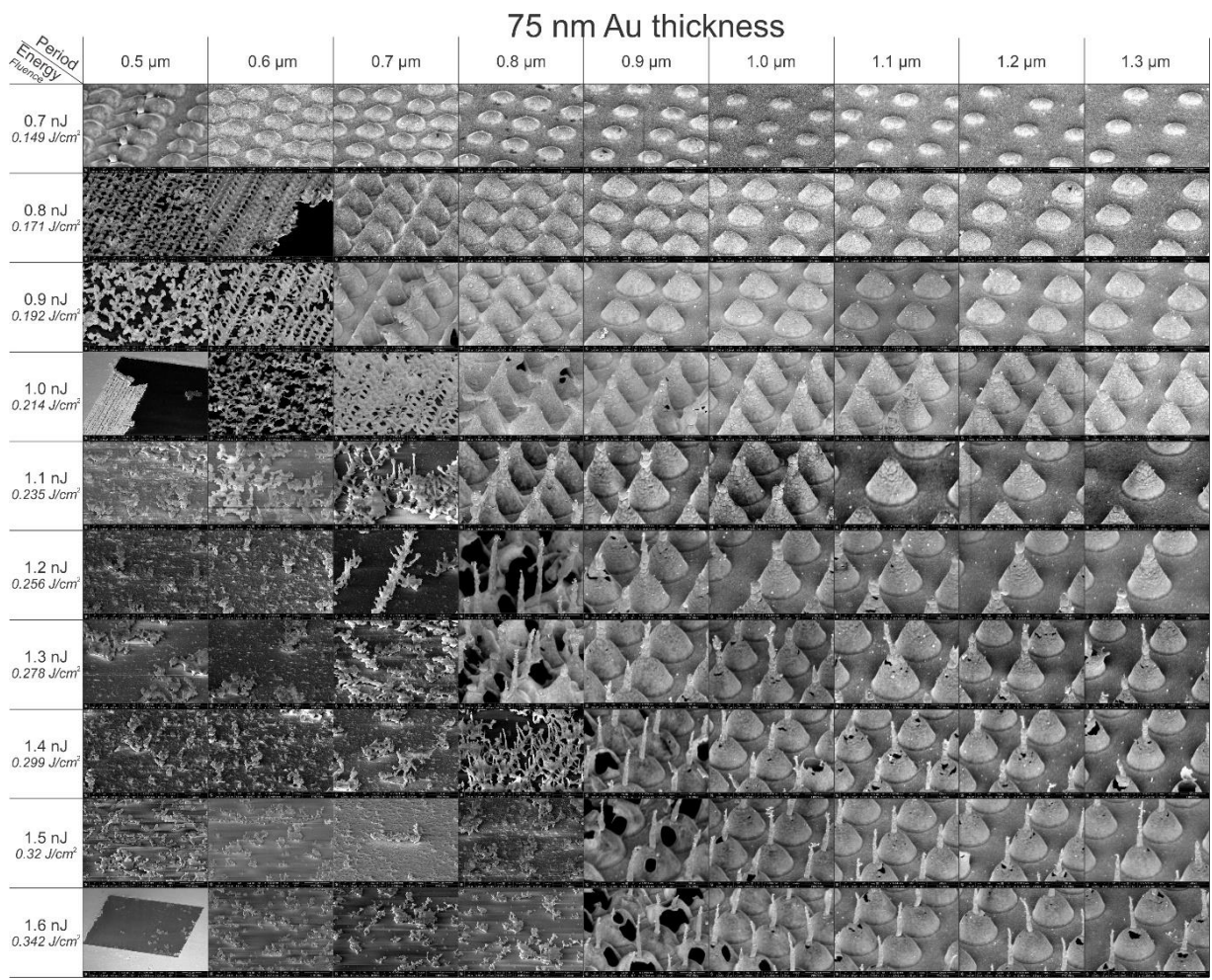

**S3.** Morphology dependence on period and energy/fluence map for 75 nm thickness Au.

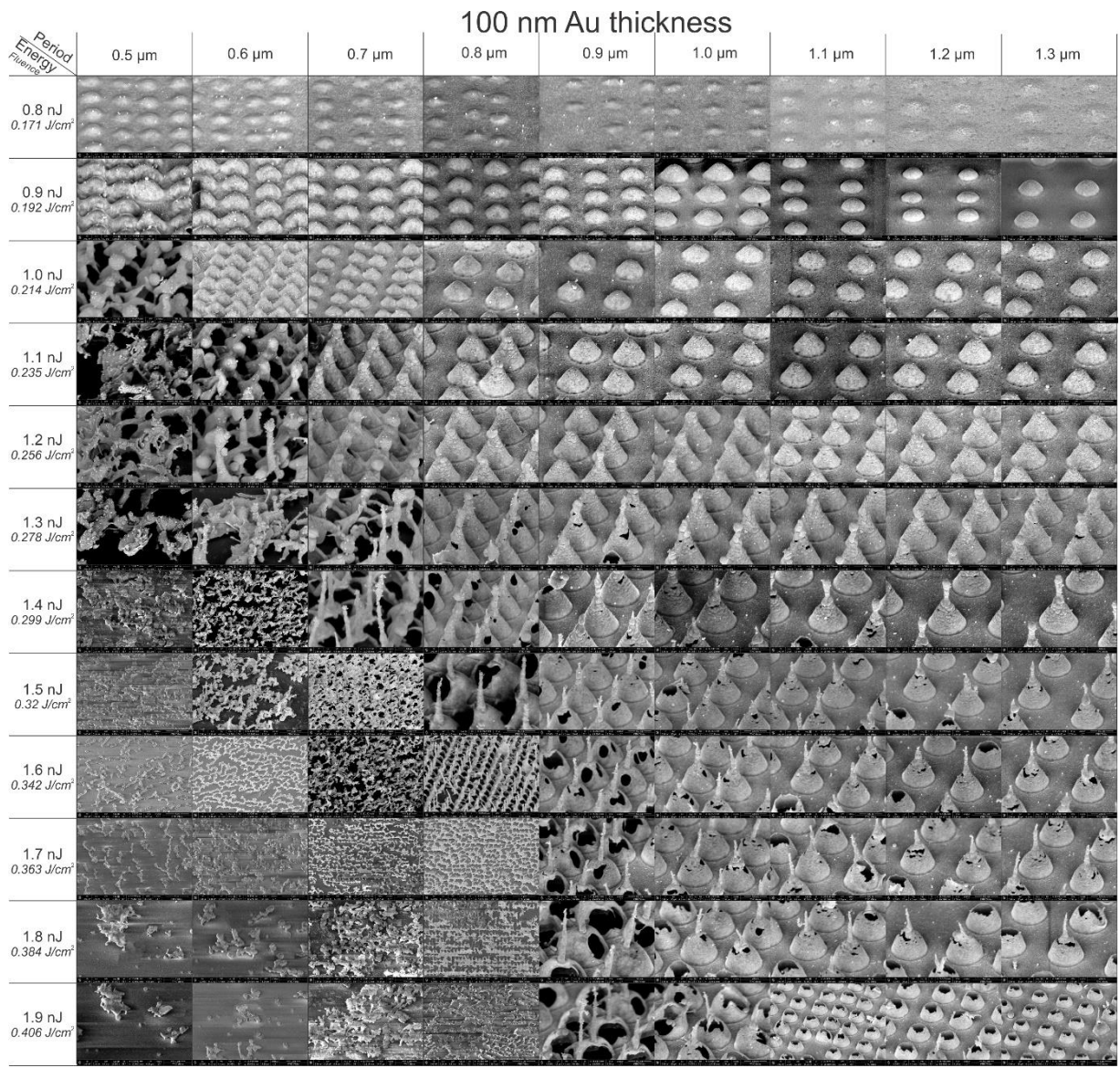

**S4.** Morphology dependence on period and energy/fluence map for 100 nm thickness Au.